\documentclass[12pt]{article}
\usepackage{graphicx}

\begin{document}
\title{COMPARISONS OF TWO QUANTILE REGRESSION SMOOTHERS}
\author{Rand R. Wilcox \\
Dept of Psychology \\
University of Southern California\\
\\
}
\maketitle
\pagebreak
\begin{center} 
ABSTRACT
\end{center}

The paper compares the small-sample properties of two non-parametric quantile regression estimators. The first is based on  constrained B-spline smoothing (COBS) and the other is based on a variation and slight extension of a running interval smoother, which apparently has not been studied via simulations. 
The motivation for this paper stems from the Well Elderly 2 study, a portion of which
was aimed at understanding the association between the cortisol awakening response and two measures of stress.
 COBS indicated what appeared be an usual form of curvature. The modified running interval smoother gave a 
strikingly different estimate, which raised the issue of how it compares to COBS in terms of mean squared error and bias as well as its ability to avoid a spurious indication of curvature. R functions for applying the methods were used in conjunction with default settings for the various optional arguments.
The results indicate that the  modified running interval smoother has practical value.  Manipulation of the optional arguments might
 impact the relative merits of the two methods, but the extent to which this is the case remains unknown.

\section{Introduction}

The paper deals with the problem of estimating and plotting a regression line when the goal is to determine the conditional quantile of some random variable $Y$ given $X$.
Quantile regression methods have been studied extensively and plots of the regression line can provide a useful perspective regarding the association between two variables. 
One approach is to assume that the
conditional $q$th quantile of $Y$, given $X$,  is given by 
\begin{equation}
Y_q = \beta_0 + \beta_1X,
\end{equation}
where $\beta_0$ and $\beta_1$ are unknown parameters. For the special case where the goal is to estimate the median of $Y$, given $X$, least absolute regression
can be used, which predates least squares regression by about a half century.  A generalization aimed at dealing with any quantile 
was derived by Koenker and Bassett (1978). While the assumption of a straight regression line appears to provide a good approximation 
of the true regression line in various situations,  this is not always the case. One strategy for dealing with any possible curvature is to use some obvious parametric model. For example, add a 
quadratic terms. But generally this can be unsatisfactory, which has led to the development of nonparametric regression lines, often called smoothers 
(e.g., H\"{a}rdle, 1990;  Efromovich, 1999;  Eubank , 1999; Gy\"{o}rfi, et al., 2002).  

For the particular case where the goal is to model the conditional quantile of $Y$ given $X$,
 one way of dealing with curvature in a reasonably flexible manner is to  
use  constrained B-spline smoothing (COBS).   The many computational  details are  summarized in 
Koenker and Ng (2005); see in particular section 4 of their paper.  
The Koenker--Ng method improves on a computational method studied by He and Ng (1999) and builds upon
  results in Koenker, Ng and Portnoy (1994). 
Briefly, let $\rho_q(u)=u(q-I(u < 0))$, where the indicator function $I(u<0)=1$ if $u<0$; otherwise $I(u<0)=0$.
The goal is to estimate the $q$th  quantile of $Y$ given $X$ by finding a function $\omega(X)$ that minimizes
\begin{equation}
\sum \rho_q(Y_i-\omega(X_i))  
\end{equation}
based on the random sample $(X_1, Y_1), \ldots, (X_n, Y_n)$. 
The estimate is based  on quadratic B-splines with  the number of knots chosen via a Schwartz-type
information criterion. Here, COBS is applied via the R package cobs.

The motivation for this paper stems from the use of COBS when analyzing data from the Well Elderly 2 study (Jackson et al., 2009; Clark et al., 2012).
A  general goal was to assess the efficacy of an intervention strategy  aimed
at improving the physical and emotional health of older adults. A portion of the study dealt with understanding the association between cortisol and various measures of
stress and well-being. 
Before and  six months  following the intervention, 
participants were asked to provide, within 1 week, four saliva samples over the course of a single day, to be obtained on rising, 30 min after rising, but before taking anything by mouth, before lunch, and before dinner. 
 Extant studies (e.g.,
 Clow et al., 2004; 
 Chida \& Steptoe, 2009) indicate that measures of
  stress are associated with the
 cortisol awakening response, which is defined as the change in cortisol concentration
that occurs during the first hour after waking from sleep. CAR is taken to be the cortisol level upon awakening  minus the level of cortisol after the participants were awake for about an hour.

After intervention (with a sample size of 328), COBS indicated some seemingly unusually shaped regression lines. One of these had to do with the association between CAR and 
a measure of depressive symptoms  using the 
Center for Epidemiologic 
Studies Depressive Scale (CESD).  The CESD (Radloff, 1977) is sensitive to change in depressive 
status over time and has been successfully used to assess ethnically diverse older people (Lewinsohn et al., 1988; Foley et al., 2002). Higher scores indicate a higher level
of depressive symptoms.
Figure 1 shows the estimated regression line for males when $q=.5$. (There were 157 males.) The estimated regression line for $q=.75$ had a shape very similar to the one 
shown in Figure 1. 

\begin{figure}
\resizebox{\textwidth}{!}
{\includegraphics*[angle=0]{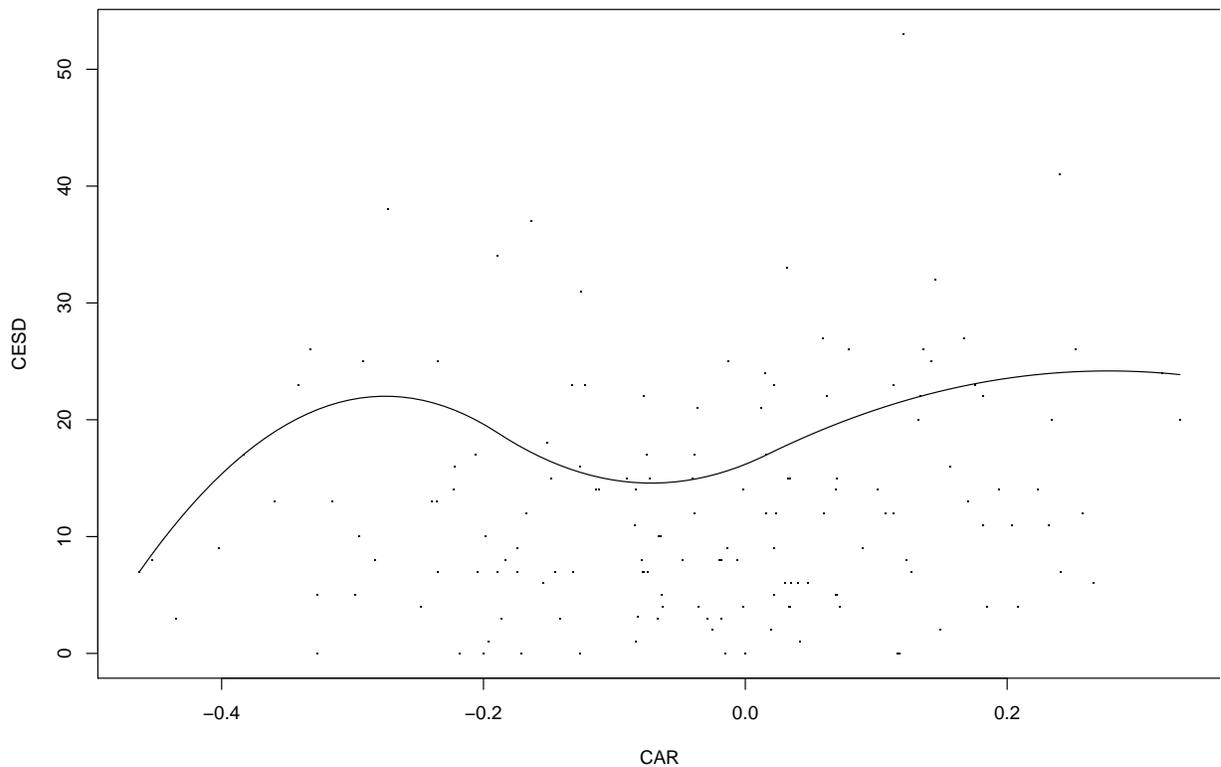}} 
\caption{COBS regression line for predicting the .5 quantile of  CESD, for males, based on  the cortisol awakening response after
intervention.}
\end{figure}

Another portion of the study dealt with the association between CAR and a measure of perceived control.  (Perceived control was measured with the
instrument in Eizenman et al. 1997. The scores ranged between 16 and 32 and consisted of a sum of Likert scales.) Now the .75 quantile regression line appears as shown in Figure 2.
Again, there was concern about the shape of the regression line. 





\begin{figure}
\resizebox{\textwidth}{!}
{\includegraphics*[angle=0]{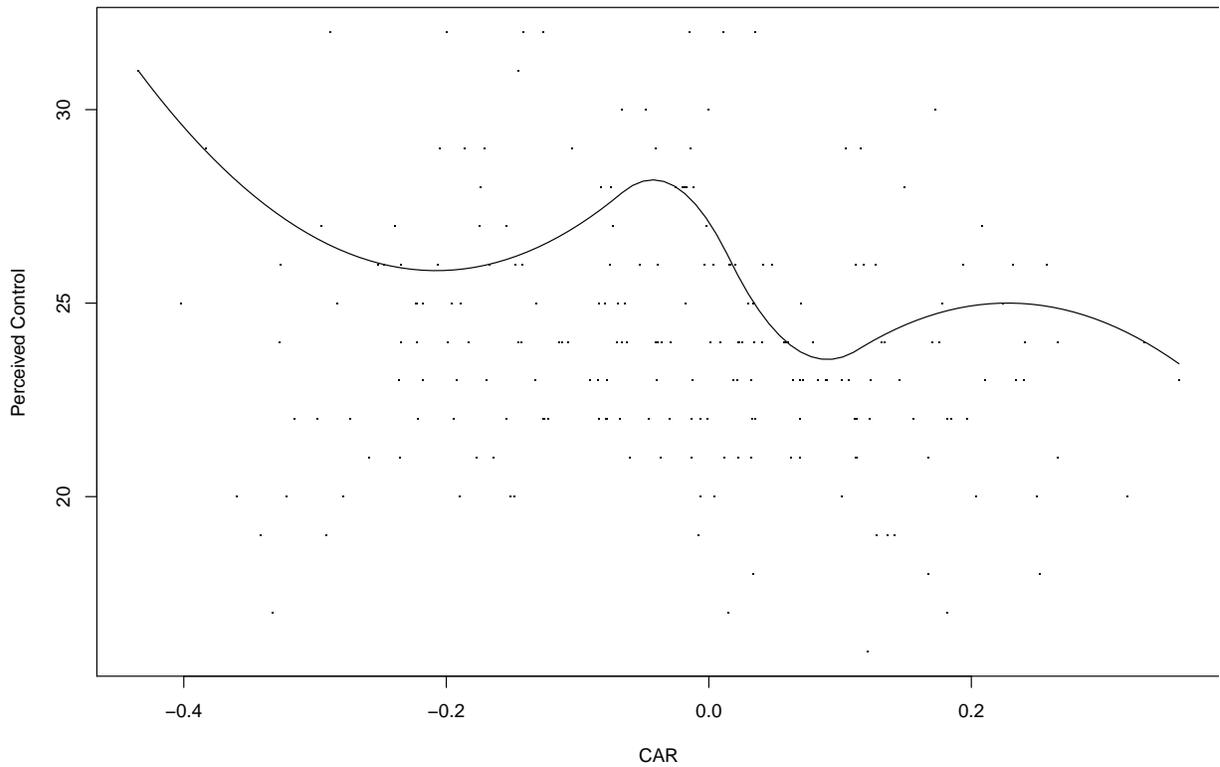}} 
\caption{COBS regression line for predicting the .75 quantile of perceived control  based on  the cortisol awakening response.}
\end{figure}


One possibility is that the regression lines in Figures 1 and 2 are a reasonable approximation of the true regression. But another possibility is that they reflect a type of curvature that
poorly approximates the true regression line. Suppose that 
\begin{equation}
Y=\beta_0+\beta_1X + \lambda(X)\epsilon
\end{equation}
where $\lambda(X)$ is some function used to model heteroscedasticity and $\epsilon$ is a random variable having mean zero and variance $\sigma^2$. 
Some preliminary simulation results suggested that if $\beta_0=0$, $\beta_1=1$, and both $X$ and $\epsilon$ have standard normal distributions, reasonably
straight regression lines are obtained using COBS. However, if $\epsilon$ has a skewed light-tailed distribution (a g-and-h distribution, details of which are described in
section 3) and if for example $\lambda(X)=|X|+1$, instances are encountered where a relatively high degree of curvature is encountered. 
An example is given in Figure 3 with $n=100$.

These results motivated consideration of an alternative quantile regression estimator. A few checks suggested that the problems just illustrated are reduced considerably, but there are
no systematic simulation results providing some sense of how this alternative estimator compares to COBS. Consequently, the goal in this paper is to compare these
estimators in terms of bias and mean squared error. Two additional criteria are used. The first is the maximum absolute error between the 
predicted and actual quantile being estimated. The other is aimed at characterizing how the estimators compare in terms of indicating a monotonic association when in fact  one exists.
This is done via Kendall's tau between the predicted and true quantiles.

It is noted that  COBS is being applied using the R package cobs in conjunction with default settings for the various arguments. The argument lambda 
alters how the regression line is estimated and might possibly improve the fit to data via  visual inspection.
But obviously this strategy is difficult to study via simulations. The alternative estimator used here is applied with an R
function (qhdsm) again using  default settings for all of the arguments. The performance
of the method is impacted by  the choice for the span (the constant $f$ in section 3). 
 The simulations reported here provide information about the relative merits of the two estimators with the 
understanding that perhaps their relative merits might be altered based on a judgmental process that goes beyond the scope of this paper. 


Section 2 provides the details of the alternative estimator. Section 3 reports simulation results comparing COBS to the alternative estimator and section 4 illustrates the difference between the
two estimators for the data used in Figures 1-3.


\begin{figure}
\resizebox{\textwidth}{!}
{\includegraphics*[angle=0]{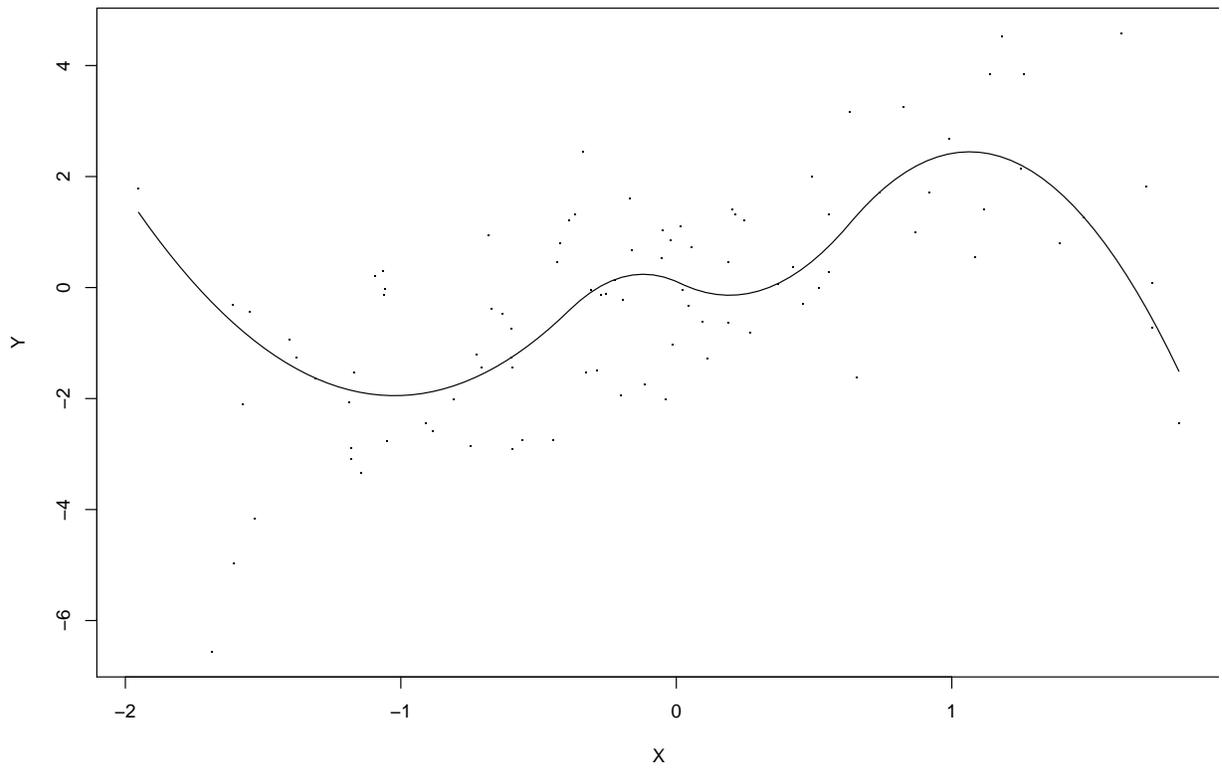}} 
\caption{COBS regression line for predicting the .5 quantile using generated data, $n=100$.}
\end{figure}

\section{Description of the Alternative Estimator}

The alternative estimator consists of a blend of two smoothers: the running interval smoother (e.g., Wilcox, 2012, section 11.5.4) and the smoother derived by
Cleveland (1979), typically known as LOWESS. The running interval has appeal because it is readily adapted to any robust estimator. In particular, it is easily applied when
the goal is to estimate the conditional quantile of $Y$ given $X$. However, often this smoother gives a somewhat jagged looking plot of the regression line. Primarily for aesthetic reasons,
this issue is addressed by further smoothing the regression line via LOWESS. 

The version of the running interval smoother used here is based in part on the quantile estimator derived by Harrell and Davis (1982).
The Harrell--Davis  estimate of the $q$th quantile uses a 
 weighted average of all the order statistics. 
 Let $Z_1, \ldots, Z_n$ be a random sample,
 let $U$ be a random variable having a beta distribution with parameters
$a=(n+1)q$ and $b=(n+1)(1-q)$, and let
 \[w_i = P\left(\frac{i-1}{n} \le U \le \frac{i}{n}\right).\]
The estimate of the $q$th quantile is
\begin{equation}
\hat{\theta}_{q} = \sum_{i=1}^n w_i Z_{(i)}, \label{hdest}
\end{equation}
where $Z_{(1)} \le \cdots   \le Z_{(n)}$ are the $Z_1, \ldots, Z_n$ written in ascending order.
Here the focus is on estimating the median and the .75 quantile. That is, $q=.5$ and .75 are used. 

In terms of its standard error, Sfakianakis and Verginis (2006) show that the Harrell--Davis estimator competes well with alternative 
estimators that again use a weighted average of all the order statistics, but there are exceptions.
(Sfakianakis and Verginis derived alternative estimators that have advantages over the Harrell--Davis in some situations. But 
when sampling from heavy-tailed distributions, the standard error of their estimators can be substantially larger than the standard error of $\hat{\theta}_{q}$.) Comparisons with other quantile
 estimators
are reported by Parrish (1990),
Sheather and Marron (1990),
as well as Dielman, Lowry and Pfaffenberger (1994). 
 The only certainty is that no single estimator dominates in terms
 of efficiency. For example, the Harrell--Davis estimator has a smaller standard error than the usual sample median when sampling from a normal
 distribution or a distribution that has relatively light tails, but for sufficiently heavy-tailed distributions, the reverse is true (Wilcox, 2012, p. 87).

The running interval smoother is applied as follows. 
Let $(X_1, Y_1), \ldots, (X_n, Y_n)$ be a random sample from some unknown bivariate distribution and let 
 $f$ be some constant to be determined.  Then the point $x$
is said to be close to $X_i$ if
\[|X_i - x| \le f \times {\rm MADN},\]
where MADN is MAD/.6745 and MAD is the median  of $|X_1-M|, \ldots, |X_n-M|$, where $M$ is the usual sample median based on $X_1, \ldots, X_n$.
For  normal distributions, MADN estimates the standard deviation, in which case 
$x$ is close to $X_i$ if
$x$ is within $f$ standard deviations of $X_i$.
Let
\[N(X_i)=\{j:|X_j-X_i| \le f \times {\rm MADN}\}.\]
That is, $N(X_i)$ indexes the set of all $X_j$ values
that are  close to $X_i$.
Let $\hat{\theta}_i$ be the Harrell--Davis estimate  based on
 the $Y_j$ values such that $j \in N(x_i)$.
To get a graphical representation of the regression line,
compute
$\hat{\theta}_i$, the estimated value of
$Y$ given that $X=X_i$, $i=1, \ldots, n$, and then plot the
points $(X_1, \hat{\theta}_1), \ldots, (X_n, \hat{\theta}_n)$ .
Typically $f=.8$ or 1 gives good results, but of course exceptions are encountered. Here, $f=.8$ is assumed unless stated otherwise. 

But as previously indicated, the plot produced by the running interval smoother can be a bit ragged. 
Consequently, the 
initial smoothed was smoothed again by proceeding as follows.   
Given $X_j$, let 
\[\delta_i=|X_i - X_j|, i=1, \dots, n.\]
Sort the $\delta_i$ values and retain the
$\xi n$ pairs of points that have the smallest $\delta_i$ values, where
$\xi$ is a number between 0 and 1 and plays the role of a span. Here, $\xi=.75$ is used. Let
$\delta_m$ be the largest $\delta_i$ value among the retained points. Let
\[Q_i=\frac{|X_j-X_i|}{\delta_m},\]
and if $0 \le Q_i <1$, set
\[w_i =  (1-Q_i^3)^3,\]
otherwise set
\[w_i = 0.\]
Next, use weighted least squares to predict $\hat{\theta}_j$ corresponding to $X_j$ using the
$w_i$ values as weights.
That is, determine the values $b_1$ and $b_0$ that minimize
\[\sum w_i(\hat{\theta}_i  - b_0 - b_1 X_i)^2\]
and estimate $\hat{\theta}_j$ with
$\tilde{{\theta}_j} = b_0 + b_1X_j$.  The final plot of the quantile regression  is taken to be the line connecting the
points $(X_j, \, \tilde{\theta}_j)$ ($j=1, \ldots, n$). This will be called method R henceforth.

\section{Simulation Results}

Simulations were used to compare the small-sample properties of COBS and the modified running interval smoother based on $K=4000$ replications and sample size $n=50$. 
The data were generated according to the model
\begin{equation}
Y=X+\lambda(X)\epsilon \label{sim.mod}
\end{equation}
where  $X$ is taken to have a standard normal distribution and $\epsilon$ has one of four distributions: 
normal, symmetric and heavy-tailed, asymmetric and light-tailed, and
asymmetric and heavy-tailed. 
More precisely, the distribution for the error term was taken to be one of four g-and-h distributions
(Hoaglin, 1985) that contain the standard  normal distribution as a special case.
If $Z$ has a standard normal distribution, then
\[W = \left\{ \begin{array}{ll}
 \frac{{\rm exp}(gZ)-1}{g} {\rm exp}(hZ^2/2), & \mbox{if $g>0$}\\
  Z{\rm exp}(hZ^2/2), & \mbox{if $g=0$}
   \end{array} \right. \]
has a g-and-h distribution where $g$ and $h$ are parameters that
determine the first four moments. 
The four distributions used here were the standard normal ($g=h=0.0$), a
symmetric heavy-tailed distribution ($h=0.2$, $g=0.0$), an asymmetric
distribution with
relatively light tails ($h=0.0$, $g=0.2$), and an asymmetric distribution with
heavy tails ($g=h=0.2$).
Table 1 shows the skewness ($\kappa_1$) and kurtosis
($\kappa_2$)
for each distribution. Additional properties of the g-and-h distribution
are summarized by Hoaglin (1985).

\begin{table}
\caption{Some properties of the g-and-h distribution.}
\centering
\begin{tabular}{ccrr} \hline
g & h &  $\kappa_1$ & $\kappa_2$\\ \hline
0.0 & 0.0  & 0.00 & 3.0\\
0.0 & 0.2 & 0.00 & 21.46\\
0.2 & 0.0  & 0.61 & 3.68\\
0.2 & 0.2 & 2.81 & 155.98\\ \hline
\end{tabular}
\end{table}

Three choices for $\lambda$ were considered: $\lambda\equiv 1$,
 $\lambda=|X|+1$, and $\lambda=1/(|X|+1)$. These three choices are called VP 1, 2, and 3
henceforth.

Note that based on how the data are generated, as indicated by (\ref{sim.mod}),  ideally a smoother should indicate a monotonic increasing association between $X_1, \ldots, X_n$ and $\breve{\theta}_{q1}, 
\ldots, \breve{\theta}_{qn}$ where  $\breve{\theta}_{qi}$ is the estimate of  the $q$th quantile of $Y$ given that $X=X_i$ based on
either COBS or method R. The
 degree to which this goal was accomplished was measured with Kendall's tau.

Details about the four criteria used to compare COBS and method R are as follows.
The first criterion was
mean squared error, which was estimated with
\begin{equation}
\frac{1}{nK} \sum_{k=1}^K \sum_{i=1}^n  (\theta_{qik} - \breve{\theta}_{qik})^2,
\end{equation}
where now, for the $k$th replication, $\theta_{qik}$ is the true conditional $q$th quantile of $Y$ given $X=X_i$ and again $\breve{\theta}_{qik}$ is the estimate of $\theta_{qik}$ based on
either COBS or method R.  
Bias was estimated with 
\begin{equation}
\frac{1}{nK} \sum_{k=1}^K \sum_{i=1}^n  \theta_{qik} - \breve{\theta}_{qik}.
\end{equation}
The third criterion was the mean maximum absolute error:
\begin{equation}
\frac{1}{K} \sum_{k=1}^K  {\rm max} \{  |\theta_{q1k} - \breve{\theta}_{q1k}|, \ldots, |\theta_{qnk} - \breve{\theta}_{qnk}|\}
\end{equation}
The fourth criterion was
\begin{equation}
\frac{1}{K} \sum_{k=1}^K  \tau_k,
\end{equation}
where for the $k$th replication, $\tau_k$ is Kendall's tau between $X_1, \ldots, X_n$ and $\breve{\theta}_{q1}, 
\ldots, \breve{\theta}_{qn}$.

It is  noted that the $\theta_{qik}$ values are readily determined because the transformation used to 
generate observations from a g-and-h distribution is monotonic and 
quantiles are location and scale equivariant.

Simulation results are reported in Table 2 where RMSE is the mean squared error of COBS divided by the mean squared error of method R and RMAX is the 
maximum absolute value of COBS divided by the 
maximum absolute value of based on method R.  As can be seen, generally method R competes well with COBS in terms of RMSE and RMAX, but neither method dominates. 
For $q=.5$, R is uniformly better
in terms of RMSE, but for $q=.75$ and VP 3 COBS performs better than R. As for RMAX, R performs best for VP 1 and 2, while for VP 3 the reverse is true. Bias for both methods is typically low with
COBS seeming to have an advantage over method R. In terms of $\tau$, method R dominates. That is, the simulations indicate that method R is better at avoiding an indication of curvature that
does not reflect the true regression line, as was the case in Figure 3.


\begin{table}
\center
\caption{Simulation Results}
\begin{tabular}{ccc    rrr rrr}
&&&&& \multicolumn{2}{c}{BIAS} & \multicolumn{2}{c}{$\tau$}\\
$g$ & $h$ & VP & RMSE & RMAX &   \multicolumn{1}{c}{COBS} &  \multicolumn{1}{c}{R}  &  \multicolumn{1}{c}{COBS} &  \multicolumn{1}{c}{R}  \\ 
   
    \multicolumn{9}{c}{$q=.5$}\\ \hline
   
0.0 & 0.0 &  1 & 1.284 & 1.293 & 0.002 & 0.002 & 0.957 & 0.997 \\
0.0 & 0.0      &  2 & 1.405 & 2.051  & $ -0.002$ & $ -0.002$ & 0.773 & 0.927\\
0.0  & 0.0       &  3 & 1.281 & 0.726  &  $ -0.001$ & $ -0.001$  &  0.989 &  1.000\\

0.0 & 0.2 &  1 & 1.160 & 1.333   &  $-0.002$  & $-0.002$  & 0.955 & 0.994 \\
0.0 & 0.2 &   2 & 1.395 & 2.076  &  0.005       & 0.009   & 0.794 & 0.917 \\
0.0 & 0.2 &   3 & 1.104 & 0.753  &  $-0.003$ & $-0.002$  &0.991 & 1.000\\ 

0.2 & 0.0 &  1 & 1.247 &  1.292  & 0.015 &  0.031 & 0.954 & 0.996 \\
0.2 & 0.0 &  2 & 1.400 &  2.048  & 0.023 &  0.035 & 0.786 & 0.930\\
0.2 & 0.0 &  3 & 1.220 & 0.732  & 0.004  &  0.022 & 0.989 & 1.000\\

0.2 & 0.2&  1 &   1.178 &  1.384 & 0.014 &  0.027 &  0.956 & 0.993\\
0.2 & 0.2&  2 & 1.455 &  2.155 & 0.034 &  0.042 &  0.794 & 0.914 \\
0.2 & 0.2 & 3 & 1.040 &  0.765 &  0.005 &  0.023 &  0.990 & 1.000\\  

 \multicolumn{9}{c}{$q=.75$}\\ \hline

0.0 & 0.0 &  1 & 1.046 &  1.375 & $-0.017$ &  0.077 &  0.938 &  0.994\\
0.0 & 0.0 & 2 &  1.328 &  2.020 &  $-0.052$ &   0.027 & 0.709 & 0.862\\ 
0.0 & 0.0 & 3  & 0.644 &   0.807 &  $-0.014$ &   0.105 & 0.973 & 0.998\\

0.0 & 0.2 &  1 & 0.847 & 1.459 &  0.010 &  0.137 & 0.911 & 0.978\\
0.0 & 0.2 &  2 & 1.124 & 2.140  &  0.022 &  0.136 & 0.666 & 0.794 \\
0.0 & 0.2 &  3 & 0.544 &  0.858 &  0.001 &  0.145 & 0.969 & 0.995\\

0.2 & 0.0 &  1 & 0.964 &  1.423 &  0.000 &  0.110 & 0.907 & 0.985\\
0.2 & 0.0 &  2 & 1.284 &  2.057 &  $-0.026$ &   0.074 & 0.655 & 0.803\\
0.2 & 0.0 &  3 & 0.642 &  0.860 & $-0.006$  &  0.126  & 0.962 & 0.997\\

0.2 & 0.2 &  1 & 0.849 & 1.505 &  0.042 &  0.181 & 0.880 & 0.953\\
0.2 & 0.2 &  2 & 1.195 & 2.214 &  0.084  & 0.202 & 0.614 & 0.740\\
0.2 & 0.2 &  3 & 0.552 & 0.945 &  0.013 &  0.167 & 0.955 & 0.993\\

\hline
 \multicolumn{9}{l}{R= modified running interval smoother}

\end{tabular}
\end{table}

\section{Some Illustrations}
The data in Figures 1-3 are used to illustrate method R. The left panel of Figure 4 shows the .5 quantile regression line for CAR and CESD. Notice that for CAR positive 
(cortisol decreases after awakening), the plot suggests a positive association with depressive symptoms, which is consistent with Figure 1. But for CAR negative, method R suggests that
there is little or no association with CESD and clearly provides a different sense regarding the nature of the association. A criticism might be that if method R were  to use a  smaller choice for the span, 
perhaps an association similar to Figure 1 would be revealed. But even with a span $f=.5$, the plot of the regression line is very similar to the one shown in Figure 4. 

The right panel of Figure 4 shows the .75 quantile regression line for predicting 
  perceived control based on CAR, which differs in an obvious way from the regression
  line based on COBS shown in Figure 2. 
  Figure 4 indicates that
   there is little or no indication of an association with CAR when CAR is negative, but for CAR positive, a negative association is indicated.
  The only point is that the choice between COBS and method R can make a substantial difference.

Figure 5 shows the .5 quantile regression line based on the data used in Figure 3. In contrast to COBS, method R provides a very good approximation of the true regression line.
Again, this only illustrates the extent to which the two methods can give strikingly different results. As is evident, in this particularly case, method R provides a much 
more accurate indication of the true regression line. 

\begin{figure}
\resizebox{\textwidth}{!}
{\includegraphics*[angle=0]{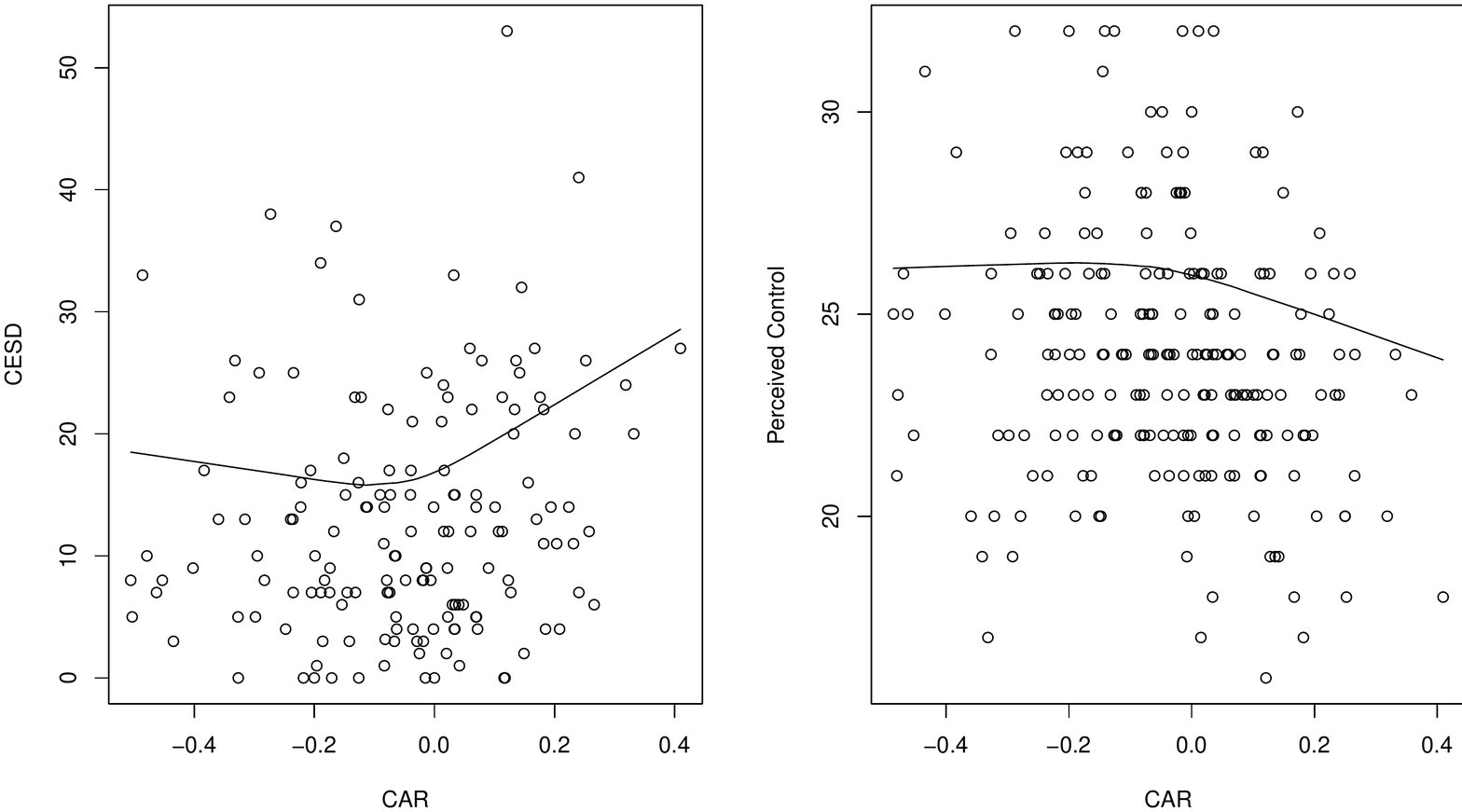}} 
\caption{The  quantile regression lines using method R and the data in Figures 1 and 2.}
\end{figure}



\begin{figure}
\resizebox{\textwidth}{!}
{\includegraphics*[angle=0]{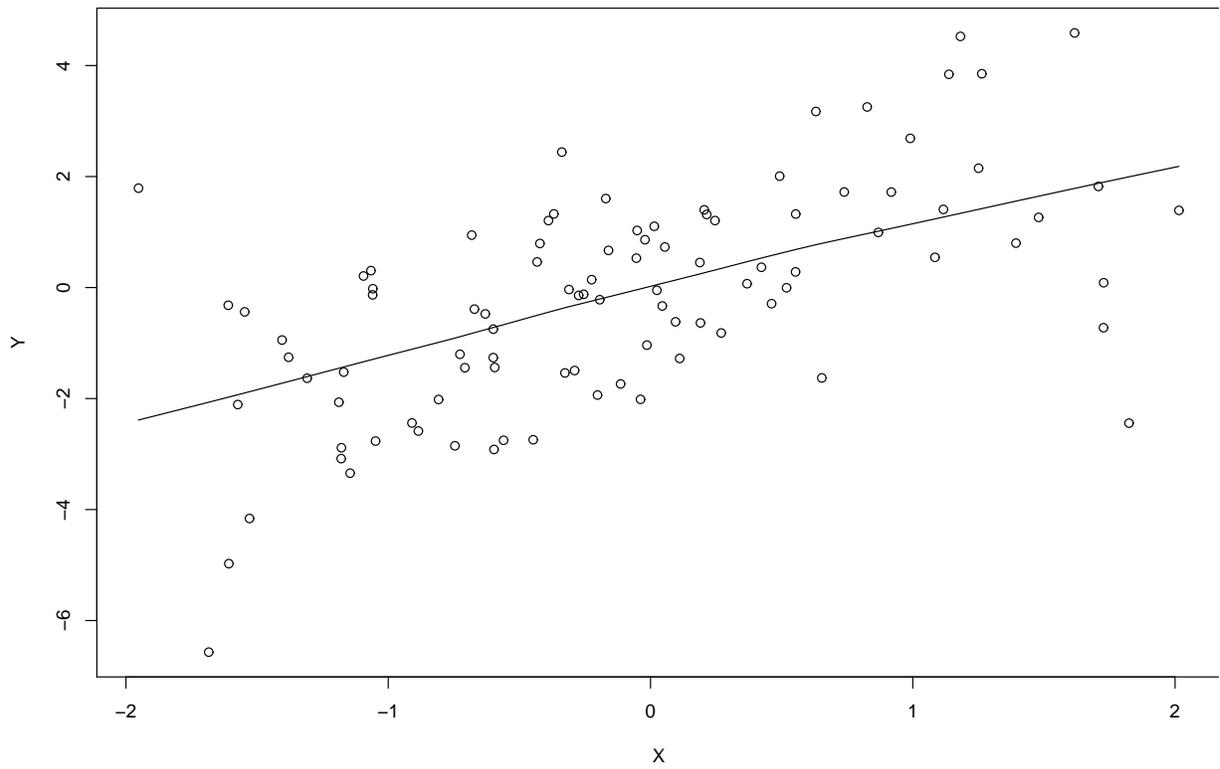}} 
\caption{The  quantile regression line using method R and the data in Figure 3}
\end{figure}

\section{Concluding Remarks}

For the situations considered in the simulations,
method R does not dominate COBS based on the four criteria used here. COBS seems to have an advantage in terms
of minimizing bias. But otherwise method R competes well with COBS, particularly in terms of Kendall's tau, which
suggests that typically method R is better able to avoid an indication of spurious curvature. Moreover,
the illustrations demonstrate that the choice between the two methods can make a substantial difference even with a sample size of $n=328$. 
So in summary, method R
would seem to deserve  serious consideration.

Another possible appeal of method R is that it is readily extended to the situation where there is more than one independent variable. That is,
 a generalization
of the running interval smooth already exists (e.g., Wilcox, 2012). Moreover, additional  smoothing can be accomplished, if desired, using the smoother derived by   Cleveland and Devlin (1988), which generalizes the technique derived by Cleveland (1979). Evidently, a generalization of COBS to more than one independent variable has not been derived. 

Finally, an R function for applying method R, called qhdsm, is available in the Forge R package WRS.
A version is also stored on the first author's web page (in the file labeled Rallfun-v26.)  
.

\begin{center}
References
\end{center}

Chida, Y. \& Steptoe, A. (2009). Cortisol awakening response and psychosocial factors: A  systematic review and meta-analysis.
 {\em Biological Psychology, 80}, 265--278.

Clark, F., Jackson, J., Carlson, M., et al. (2012). Effectiveness of a lifestyle intervention in promoting
the well-being of independently living older people:
results of the Well Elderly 2 Randomise Controlled Trial.  {\em Journal of Epidemiology and Community Health, 66},
 782--790. doi:10.1136/jech.2009.099754
 
 Cleveland, W. S. (1979). Robust locally weighted regression and smoothing
 scatterplots. {\em Journal of the American Statistical Association, 74},
 829--836.
 
 Cleveland, W.S., and Devlin, S.J., (1988). Locally-weighted Regression:   An Approach to Regression Analysis by Local Fitting.
{\em Journal of the American Statistical Association, 83},  596--610.
 
 Clow, A., Thorn, L., Evans, P. \&  Hucklebridge, F. (2004). The awakening cortisol response: Methodological issues and significance.
{\em Stress, 7}, 29--37.
 
 Dielman, T., Lowry, C. \& Pfaffenberger, R. (1994). A comparison of quantile estimators. {\em Communications in Statistics--Simulation and}
{\em Computation, 23}, 355-371.
 
Jackson, J., Mandel, D., Blanchard, J., Carlson, M., Cherry, B., Azen, S., Chou, C.-P.,  Jordan-Marsh, M., Forman, T., White, B., Granger, D., Knight, B., \& Clark, F. (2009). Confronting challenges in intervention research with ethnically diverse older adults: the USC Well Elderly II trial. {\em Clinical Trials, 6},  90--101.

Efromovich, S. (1999). {\em Nonparametric Curve Estimation: Methods, Theory and Applications}. New York: Springer-Verlag.

Eizenman, D. R., Nesselroade, J. R., Featherman, D. L. \& Rowe, J. W. (1997). Intraindividual variability in perceived control in an older sample: The MacArthur Successful Aging Studies.
{\em Psychology and Aging, 12}, 489--502.

Eubank, R. L. (1999). {\em Nonparametric Regression and Spline Smoothing}. New York: Marcel Dekker.

Foley K., Reed P., Mutran E., Devellis, R. F. (2002). Measurement adequacy of the CESD  among a sample of older African Americans. {\em Psychiatric Research, 109}, 61--9.

Gy\"orfi, L., Kohler, M., Krzyzk, A., \&  Walk, H. (2002).  {\em A Distribution-Free Theory of Nonparametric Regression}. New York:  Springer Verlag.

H\"{a}rdle, W. (1990). {\em Applied Nonparametric Regression}. Econometric Society Monographs No. 19, Cambridge, UK:  Cambridge University Press.

Harrell, F. E. \& Davis, C. E. (1982). A new distribution-free quantile estimator. {\em Biometrika, 69}, 635--640.

He, X. \& Ng, P. (1999). COBS: Qualitative constrained smoothing via linear programming.
 {\em Computational Statistics, 14}, 315--337.

Hoaglin, D. C. (1985). Summarizing shape numerically: The g-and-h distribution.
 In D. Hoaglin, F. Mosteller \& J. Tukey (Eds.) {\em Exploring Data Tables}
 {\em Trends and Shapes}. New York: Wiley.

Koenker, R. \& Bassett, G. (1978). Regression quantiles. {\em Econometrika, 46},
 33--50.

Koenker, R. \& Ng, P. (2005). Inequality Constrained Quantile Regression.
  {\em Sankhya, The Indian Journal of Statistics, 67},  418-440.

Koenker, R., Ng, P. \& Portnoy, S. (1994). Quantile smoothing splines. {\em Biometrika, 81},
 673--680.

Lewinsohn, P.M., Hoberman, H. M., \& Rosenbaum M. (1988). A prospective study of risk factors  for unipolar depression. {\em Journal of  Abnormal Psychology, 97}, 251--64.

Parrish, R. S. (1990). Comparison of quantile estimators in normal sampling.
 {\em Biometrics, 46}, 247--257.

Radloff, L.  (1977). The CESD scale: a self report depression scale for research in the general population. {\em Applied Psychological Measurement, 1}, 385--401.

Sfakianakis, M. E. \& Verginis, D. G. (2006). A new family of nonparametric quantile estimators.
{\em Communications in Statistics--Simulation and Computation, 37}, 337--345.

Sheather, S. J. \& Marron, J. S. (1990). Kernel quantile estimators. {\em Journal}
 {\em of the American Statistical Association, 85}, 410--416.

Wilcox, R. R. (2012). {\em Introduction to Robust Estimation and
Hypothesis Testing}, 3rd Ed. New York: Elsevier.

\end{document}